\documentclass[prb,twocolumn,showpacs,floats,amssymb,superscriptaddress]{revtex4}

\usepackage{graphics}
\usepackage{epsfig}
\usepackage{dcolumn}
\usepackage{amsmath}

\begin{document}
\title{Negative photoconductance in a biased multiquantum well with filter barriers}

\author{Anibal T. Bezerra}
\email{anibal@df.ufscar.br}
\affiliation{Departamento de F\' \i sica, Universidade Federal de S\~ ao Carlos,
13565-905, S\~ ao Carlos, SP, Brazil}
\affiliation{DISSE - Instituto Nacional de Ci\^ encia e Tecnologia de Nanodispositivos
Semicondutores, CNPq/MCT, Brazil}
\author{Marcelo Z. Maialle}
\affiliation{Faculdade
de Ci\^ encias Aplicadas, Universidade Estadual de Campinas,
13484-350, Limeira, SP, Brazil}
\affiliation{DISSE - Instituto Nacional de Ci\^ encia e Tecnologia de Nanodispositivos
Semicondutores, CNPq/MCT, Brazil}
\author{Marcos H. Degani}
\affiliation{Faculdade
de Ci\^ encias Aplicadas, Universidade Estadual de Campinas,
13484-350, Limeira, SP, Brazil}
\affiliation{DISSE - Instituto Nacional de Ci\^ encia e Tecnologia de Nanodispositivos
Semicondutores, CNPq/MCT, Brazil}
\author{Paulo F. Farinas}
\affiliation{Departamento de F\' \i sica, Universidade Federal de S\~ ao Carlos,
13565-905, S\~ ao Carlos, SP, Brazil}
\affiliation{DISSE - Instituto Nacional de Ci\^ encia e Tecnologia de Nanodispositivos
Semicondutores, CNPq/MCT, Brazil}
\author{Nelson Studart}
\affiliation{Departamento de F\' \i sica, Universidade Federal de S\~ ao Carlos,
13565-905, S\~ ao Carlos, SP, Brazil}
\affiliation{DISSE - Instituto Nacional de Ci\^ encia e Tecnologia de Nanodispositivos
Semicondutores, CNPq/MCT, Brazil}

\date{\today}

\begin{abstract}
In this paper the photon-assisted electron motion in a multiquantum well (MQW) semiconductor heterostructure in the presence of an electric field is investigated. The time-dependent Schr\"odinger equation is solved by using the split-operator technique to determine the photocurrent generated by the electron movement through the biased MQW system. An analysis of the energy shifts in the photocurrent spectra reveals interesting features coming from the contributions of localized and extended states on the MQW system. The photocurrent signal is found to increase for certain values of electric field, leading to the analogue of the negative-conductance in resonant tunneling diodes. The origin of this enhancement is traced to the mixing of localized states in the QWs with those in the continuum. This mixing appears as anticrossings between the localized and extended states and the enhanced photocurrent can be related to the dynamically induced Landau-Zener-St\"uckelberg-Majorana transition between two levels at the anticrossing.
\end{abstract}


\maketitle
\newpage

\section{Introduction}

It is well known that the application of external electric fields to quantum objects leads to intriguing phenomena like, among others, the quantum-confined Stark effect which describes the effect of an external electric field upon the optical spectrum of a quantum well (QW).~\cite{Mil84} Another striking phenomenon is the carrier localization in single miniband of a superlattice under strong electric field, the Wanier-Stark ladder.~\cite{Mend88,Cury87,Deg91} Nowadays, fuelled by technological innovation, old physical concepts and phenomena found in semiconductor heterostrutures have been revived and broad avenues are still open for designing new devices ranging from quantum cascades lasers,~\cite{Fais96,Gmac98} and, for the sake of our interest, to the infrared (IR) photodetectors based on intrasubband tunnelling between states localized in quantum wells~\cite{Lev93,Dre95,Gan98,Deg11} or/and quantum dots (QD),~\cite{Lim05,Souza08,Deg11a,Alv12,Kim13} and superlattices.~\cite{Kea95,Zeu96a,Zeu96b} Even though most of these devices are now commercially available, there are presently a lot of unresolved questions concerning the optimal design and even the basic physical mechanisms underlying the behavior of these systems. Laterally, the gradual improvement of computational approaches has been yielding investigations of, e.g., out-of-equilibrium transport of charge, and non-linear phenomena such as the role of multiple-photon absorptions.~\cite{Deg10}

In the present work a theoretical approach based on simple model 
suited for charge transport calculations is adopted in the investigation
of a structure composed by alternating wells and barriers generating a multiple quantum well (MQW) profile, under the simultaneous application
of static and oscillating biases. The proposed structure has additional barriers inside the QW barriers, whose potential confinement exceeds the QW ones, generating a superlattice-like potential profile above the wells (for more details, see Section II). Given the length scales
of the quantum wells analyzed (in the tenths of nanometer range)
the classical oscillating field produces the same effects, from
the computational perspective, as if fully quantized photon-fields were
used. Hence, the oscillating electric field can
be viewed either as a laser beam, given the frequencies chosen,
in the THz domain, or as a very fast-oscillating
alternated field, driving the structure out of equilibrium.
We work within the latter view but adopt the former view
by calling the resulting charge motion throughout the structure 
by ``photocurrent''.

%%%****Fig 1*****%%%%%%%%%%%%%%%%%%%%%%%%%%%
\begin{figure}
\linespread{1.0}
\includegraphics[width=8.4cm]{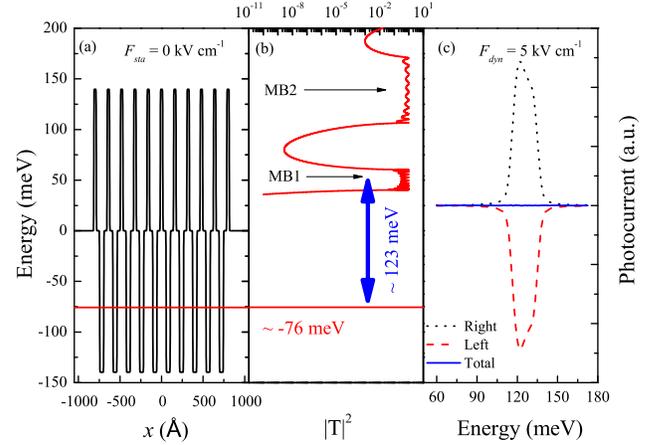}
\caption{\label{fig-potential}(color online) (a) Potential profile of the structure composed by alternating wells and barriers generating a multiple quantum well profile. The additional potential barriers inside the QW barriers generate a superlattice-like potential profile above the wells contributing for the formation of the minibands shown in (b).
(b) Transmission coefficient through the structure, showing the minibands MB1 and MB2. (c) Photocurrent for zero electric field.
The black-dotted (red-dashed) line represents the right (left) component 
of the photocurrent and the blue line represents the total photocurrent.}
\end{figure}
%%%%%%%%%%%%%%%%%%%%%%%%%%%%%%%%%%%%%%%%%%

The photocurrent generated is shown as emerging from
simple (yet usual) model assumptions. 
As expected, a series of resonances appear in the plot
of photocurrent versus photon-energy whose peaks are
understood in terms of the contributions of each
individual quantum well to the photocurrent.
Stark shifts of these resonances are described
in terms of the electronic states involved in the
photocurrent generation, and a detailed analysis revels the role
of the localized and extended states.
Among the features observed in the photocurrent spectra
is the generation of reverse (against bias)\cite{Sir93} 
photocurrents for certain photon wavelengths and low enough biases, 
which are known to appear in measurements.
The resonant photocurrent signal shows an interesting
dependence on the applied electric fields, namely an enhanced photocurrent signal
for certain values of field, leading to a behavior
which resembles the negative-conductance phenomenon known in 
resonant tunneling diodes.\cite{Chang74, Soll83}
The origin of this enhancement is traced to the mixing of localized states in the MQW
structure with the extended states in the energy continuum. This
mixing appears as anticrossings between the localized
and extended states such that the enhanced photocurrent
can be related to the Landau-Zener-St\"uckelberg-Majorana (LZSM) transition,\cite{Lan32,Zen32,Stuc32,Maj32}
that is, a transition dynamically  induced between two
levels at an anticrossing.\cite{Rib13,Rib13a}

\section{Multiple Quantum Well System and Theory}\label{secII}

Our MQW structure illustrated in the Fig.~\ref{fig-potential}(a) is formed by ten GaAs QWs of 5nm thickness and Al$_{0.15}$Ga$_{0.85}$As barriers of 11 nm. Inside each barrier of the MQW we place thin  Al$_{0.30}$Ga$_{0.70}$As filter barriers of 3 nm, generating a superlattice-like potential profile above the wells. This profile is conceived to create localized states in the MQW-continuum in such way that their coupling to the system ground state is increased, while keeping them sufficiently extended to warrant generation of photocurrent. The system is designed to have, in the absence of external biases, 10 very narrow spaced levels around -76 meV inside the MQW and a series of minibands in the continuum region above the QWs. As we can see in the Fig.~\ref{fig-potential}(b) the lowest miniband (MB1) is around +47 meV.

Figure \ref{fig-potential}(b) shows the transmission coefficient for our
structure for zero electric field. The minibands have ten distinguishable peaks, in correspondence to the ten wells of the MQW. We can observe that the energy separation between the bound states in the QWs and the
center of the MB1 is approximately 123 meV, which agrees very well with the photocurrent peak shown in
Fig.~\ref{fig-potential}(c). As a consequence of the structure symmetry, the net photocurrent is zero, at zero electric field, where the contributions of right and left current peaks cancel out. From these individual peaks, one electron is photoexcited from the QW bound state to the first miniband by a photon of $\sim$123 meV, and escapes from the filter barriers region to contribute to the photocurrent signal.

The electron excitation from the bound states in the QW is provided by the application of an oscillating electric field perpendicular to the heterostructure layers.
Also a static electric field is applied in the same direction of the oscillating field.
The Hamiltonian of the electron in the effective-mass approximation is hence given by
\begin{equation}\label{EQ01}
\widehat{H} = -\frac{\hbar^{2}}{2m^{*}}\frac{d^{2}}{dx^{2}}+V(x) 
- ex(F_{sta} - F_{dyn}\sin(\Omega t)), 
\end{equation} 
which is the one dimensional part of the full Hamiltonian with the usual assumptions of conservation of the momentum parallel
to the layers, $V(x)$ is the profile potential of the structure
sketched in the Fig.~\ref{fig-potential}(a), $m^{*}$ is the electron effective mass 
considered uniform throughout the system, and $e$ is the electron charge. In the Eq.(~\ref{EQ01}), $F_{dyn}$ and $F_{sta}$ are the intensity of the oscillating and static field, respectively. 

We use a numerical approach to determine the time evolution of the wave functions given by
\begin{equation}\label{EQ02}
\Psi(x,t+\Delta t) = e^{-i\widehat{H}\Delta t /\hbar}\Psi(x,t), 
\end{equation}
where $\Delta t$ is the time increment, $\hbar$ is the reduced Planck constant and $\widehat{H}$ 
is the system Hamiltonian within the effective-mass approximation, given by Eq.~(\ref{EQ01}).  
Since the kinetic operator, $\widehat{T}$, and potential operator, $\widehat{V}$, in the Hamiltonian 
do not commutate, the exponential operator on
the Eq.~(\ref{EQ02}) can not be performed exactly 
and some approximations need to be used. In the present work
we have used the split-operator technique~\cite{Deg10}
\begin{equation}
e^{-i(\widehat{T}+\widehat{V})\Delta t/\hbar}
= e^{-i\widehat{V}\Delta t/2\hbar}e^{-i\widehat{T}\Delta t/\hbar}e^{-i\widehat{V}\Delta t/2\hbar}
+O(\Delta t^{3}).\nonumber
\end{equation}
Thus, successive applications of this time-evolution operator evolve 
the initial wave function from $t=0$ to $t>0$, within an error of the 
order of $\Delta t^{3}$. If this procedure is realized in imaginary 
time, making $t\rightarrow -i\tau$ and setting $F_{dyn}=0$, we can 
calculate the eigenstates and eigenenergies of the MQW structure as a 
function of the electric field.~\cite{Deg10} For the present method we 
used hard-walls as boundary conditions, i.e., the wave function 
vanishes at the boundaries. To avoid reflections of the wave functions 
at the boundaries, we implemented exponential imaginary absorber 
barriers on the potential at large distances from the system's active 
region.~\cite{Mai11} 

The particle current, flowing toward both sides of the system can be computed as
\begin{equation}
J_{c}(t)=\Re\left[\frac{\hbar}{im^{*}}\Psi(x,t)^{*}
\frac{\partial\Psi(x,t)}{\partial x}\right]_{x=x_{c}},
\end{equation} 
where $\Psi(x,t)$ is the system wave function under the oscillating 
electric field and the index $c=left$ or $c=right$ represents the left and right components of the 
photocurrent (Fig.~\ref{fig-potential}(c)). The current $J_{c}$ is then integrated over time to obtain
\begin{equation}
I=\frac{e}{T_{p}}\int^{T_{p}}_{0}\left(J_{right}(t)-J_{left}(t)\right)dt,
\end{equation}
where $T_{p}$ is an upper bound time which depends on the frequency of 
the oscillating field. A detailed discussion of this technical point 
is made in Ref~\onlinecite{Mai11}. The net photocurrent is given by 
the sum of the photocurrents generated by initial states localized in 
each QW. In what follows, ohmic effects are not taken into account and 
the photocurrent is purely coherent.

\section{Results and discussion}

Hereafter we number orderly the QWs (shown in Fig.~\ref{fig-potential}(a)) from the right-hand side to the left-hand side by 
$1^{st}$ QW up to $10^{th}$ QW. As expected, when the structure is 
biased the energy level of the $1^{st}$ QW is lowered relatively to 
the $10^{st}$ QW due to the Wannier-Stark effect.\cite{Mend88, BLE88} 

\begin{figure}
\includegraphics[width=7.3cm]{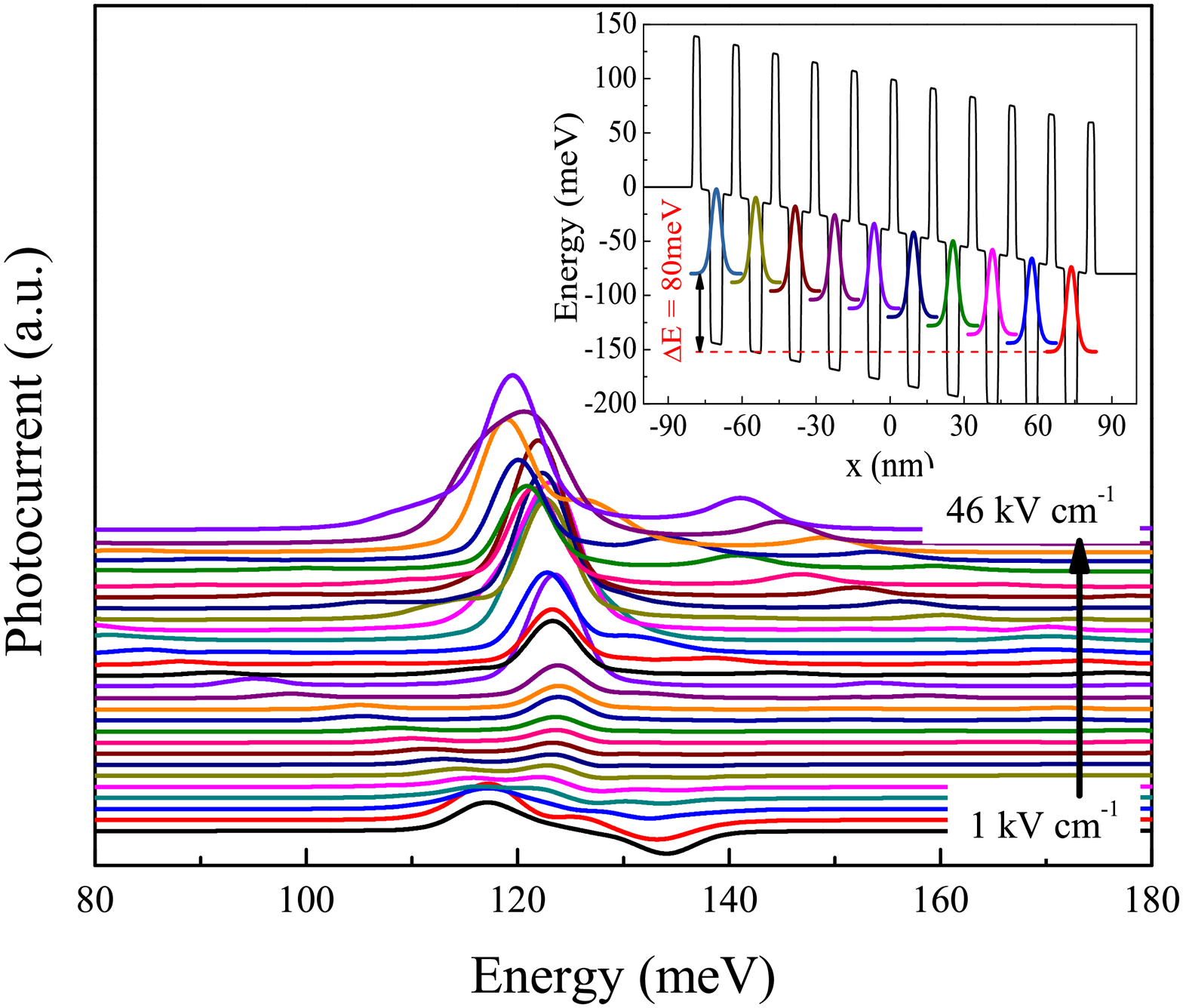}
\caption{(color online) Photocurrent spectra for different applied
electric fields. Curves are vertically shifted for better visualization.
The inset shows the potential profile and the 
ground state wave function localized
in each well for a 5 kV cm$^{-1}$ electric field.
We observe that the localized states spread over 
a energy region of $\Delta E\simeq$80 meV.
As a thumb rule the regions where
the photcurrents are constant (straight horizontal lines) can
be taken as the zero-value baseline for a vertical
scale (note that some values of the photocurrent are thus
negative).}
\label{fig-TotalCur}
\end{figure}

In Fig.~\ref{fig-TotalCur} the dependence of the photocurrent with the 
electric field is depicted for $T_{p}$=1 ps and $F_{dyn}$=5 kVcm$^{-
1}$. Due to the breaking of the potential symmetry, the states became 
localized with an energy shift of $\Delta E = edF_{sta}$, where $d$ is 
the period of the structure. As shown in the inset of the 
Fig.~\ref{fig-TotalCur}, we clearly observe the effect of the Wannier-
Stark localization as the states spread over a 80 meV energy region 
for an electric field of 5 kVcm$^{-1}$. The same occurs with the 
miniband states in the continuum.

%%%%%%%%%%%%%%%%%%%%%%%%%%%%%%%%%%%%%%%%%
%%%%%%%%%%%%%%%%%%%%%%%%%%%%%%%%%%%%%%%%%%

%%%%%%%%%%%%%%%%%%%%%%%%%%%%%%%%%%%%%%%%%%%%
\begin{figure}
\includegraphics[width=8cm]{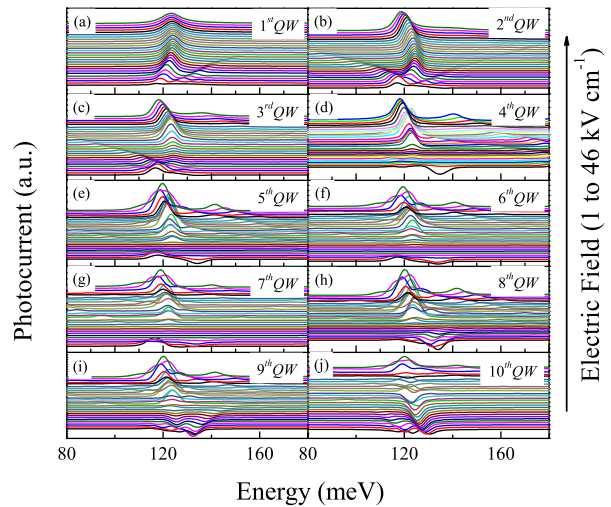}
\caption{(color online) Photocurrent spectra for different electric fields ranging from 1 to 46 kV cm$^{-1}$.
The panels (a) to (j) are related to the photocurrent of each individual QW ($1^{st}$ to $10^{th}$),
respectively. Curves were vertically shifted for better visualization and different
scales are used in each graph. As a thumb rule the regions where
the photcurrents are constant (straight horizontal lines) can
be taken as the zero-value baseline for a vertical
scale (note that some values of the photocurrent are thus
negative).}
\label{fig-PCbyQW}
\end{figure}
%%%%%%%%%%%%%%%%%%%%%%%%%%%%%%%%%%%%%%%%%%%%%

In order to better understand the total photocurrent spectra in Fig.~\ref{fig-TotalCur} we look separately to each QW photocurrent contribution as shown in Fig.~\ref{fig-PCbyQW}.
We noticed (not shown here) that for low and high bias regimes, the major contribution to the total photocurrent comes from the more external QWs. That is, mostly from the two first QW on the right-hand side and the two last on the opposite side. Then it is enough for clear understanding focus our discussion on the mentioned QWs.

In Fig.~\ref{fig-PCbyQW}(a), which represents the photocurrent of $1^{st}$ QW, 
we can see for low biases two positive peaks. 
Both peaks blueshift with increasing the electric field and the Wannier-Stark relationship $\Delta E = edF_{sta}$ is approximately obeyed. 
We associate the main photocurrent peak in Fig.~\ref{fig-PCbyQW}(a) to the transition from the ground state of the $1^{st}$ QW to the lowest state of the first miniband (MB1) located in the filter barriers region. Both states are spatially localized in the $1^{st}$ QW region, therefore enhancing the overlap of the wave functions that favors the transition and, consequently, the photocurrent signal. For simplicity, even in the cases for which there is localization we  will use the term ``miniband"    in reference to these states.
The second photocurrent peak seen in 
Fig.~\ref{fig-PCbyQW}(a) is related to the transition
between the $1^{st}$ QW bound state and the second lowest-energy state of the MB1
which is spatially localized in a neighboring $2^{nd}$ QW.
In this case, the relative energy shift between these states in the presence of bias signals the Stark ladder formation.\cite{Deg91} As the delocalized states of the miniband state becomes localized in the $2^{nd}$ QW, 
the overlap with the $1^{st}$ QW bound state decreases,
resulting in the observed decrease of the
second peak. A similar argument is valid for the photocurrent 
contribution of the $2^{nd}$ QW and shown in the Fig.~\ref{fig-PCbyQW}(b). 

We now discuss the contributions to the overall photocurrent of the $9^{st}$ QW and $10^{st}$ QW in the left-hand side of the  MQW structure with filter barriers. As we can see in Figs.~\ref{fig-PCbyQW}(i) and (j), the photocurrent contribution of these QWs presents negative values (i.e. current flowing against the electric field) for low bias which become positive with increasing the electric field. This phenomenon was already demonstrated and well-discussed by Sirtori and collaborators,~\cite{Sir93} and is related to 
the transition between the ground state and the higher energy states of MB1 which are extended over the left-hand side of the MQW structure allowing the 
electrons to conduct against the electric field direction.
By further increasing the electric field the whole structure becomes 
transparent to the photoexcited electrons favoring the action of the electric field in producing positive photocurrent. The shifts in the peak positions and the decrease in the peak intensities are also related to Stark ladder formation and to decrease of the overlap between wave functions of different QWs, respectively. 

\begin{figure}
\includegraphics[width=8.0cm]{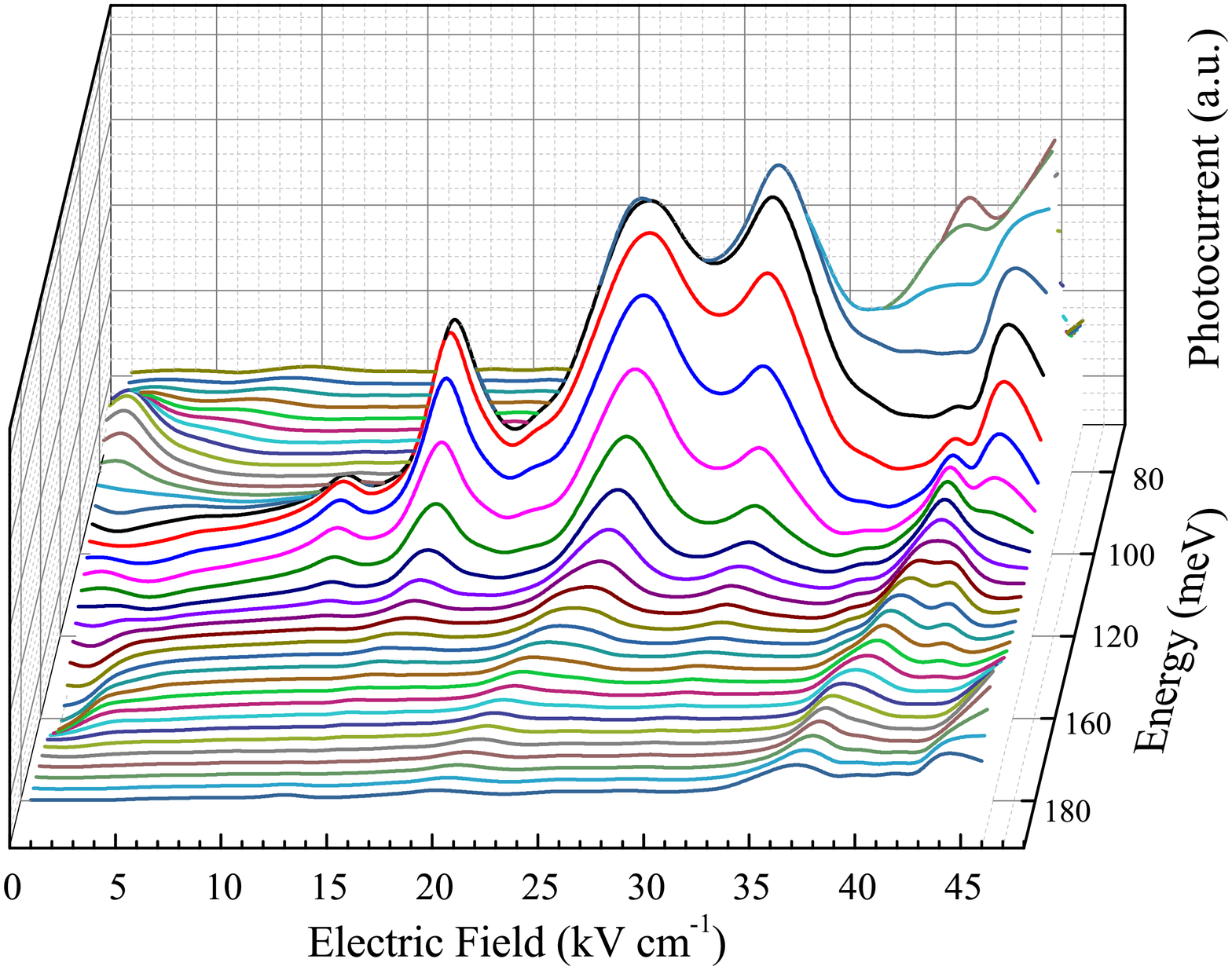}
\caption{\label{fig-IV}(color online) Photocurrent is depicted as a function of the electric field and the energy showing the peaks shifts discussed in the text.}
\end{figure}

Looking more carefully to the photocurrent dependence on bias, we see in Fig.~\ref{fig-IV} (same results of  Fig.~\ref{fig-TotalCur} in a different glance) that the intensity of the main peak at about 120 meV first decreases, but next starts to increase for certain electric field values.
The origin of the first decrease observed for intermediate biases is the Stark shifts of
the positive and negative peaks, already discussed
in Figs.~\ref{fig-PCbyQW}(a) and (j), respectively.
We again call the attention, from these figures, that the positive peak slightly blueshifts 
while the negative one redshifts with a bias increase.
As an expected consequence, in the intermediate bias region the peaks intensities partially cancel out each other contributing to diminish the total photocurrent. 

%%%%%%%%%%%%%%%%%%%%%%%%%%%%%%%%%%%%
\begin{figure}
\includegraphics[width=7.5cm]{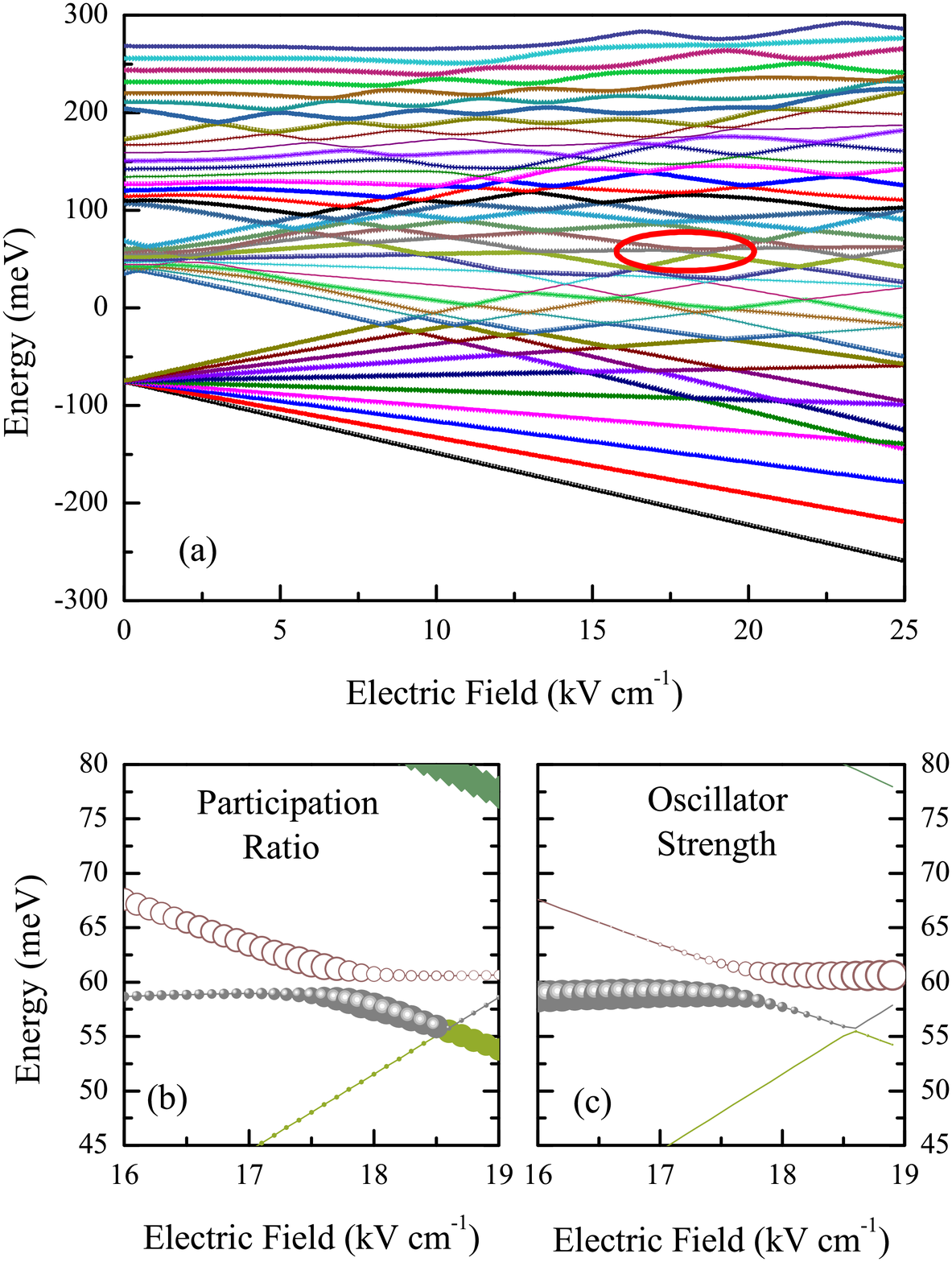}
\caption{\label{fig-states}(color online) (a) Energy spectrum as a function of the electric field.
(b) Zooming of the encircled region of the energy spectrum of (a) showing the levels anticrossing. The dot sizes are proportional to the participation ratio which means that the larger the dot more the state is delocalized. (c) Same as (b) with the dot sizes proportional to the oscillator strength.}
\end{figure}

%%%%%%%%%%%%%%%%%%%%%%%%%%%%%%%%%%%%%
The photocurrent enhancement seen in Fig.~\ref{fig-IV} for certain electric fields, around the 
excitation energy 120 meV has a different origin, namely
the mixing between extended and localized states. Because we construct our system placing a thinner filter barrier inside each barrier of the MQW structure which led to the formation of very close quasi-continuum states (miniband-like, as we call earlier) and still enough localized in the QWs region, we achieve a bias dependent coupling of the QW ground states and the extended states in the continuum (above the MQW structure). In order to make clear this state coupling, we present in  Fig.~\ref{fig-states}(a) the MQW energy spectrum as a function of the electric field. For increasing bias the miniband states Stark shift and many crossings are seen.
In Fig.~\ref{fig-states}(b) and (c) we focus on the region of the energy spectrum given by the red circle in Fig.~\ref{fig-states}(a), in which the anticrossing between the first and second minibands is clearly depicted. The dot sizes in Fig.~\ref{fig-states}(b) are
proportional to the participation ratio, 
\begin{equation}
P = \frac{1}{L}\frac{\left(\int|\Psi(x)|^{2}dx\right)^{2}}{\int|\Psi(x)|^{4}dx},\nonumber
\end{equation}
where the integration is over the entire space, $L$ is the size of 
the system, and the numerator will be one for states normalized
to unity.
$P$, as introduced by Bell and Dean~\cite{Dean70} and also 
described by other authors~\cite{Aoki83,Edwa72}, is basically a 
measure of how much extended is the state, in other words, extended 
states have larger $P$. As we can observe in Fig.~\ref{fig-states}(b), 
the crossing occurs between an initially localized state of the MB1, 
with smaller $P$ (solid gray dotted curve) and an initially extended 
state of the MB2, with greater $P$ (open brown dotted curve). During 
the crossing, both states has comparable participation ratio, changing 
their character after the crossing. Namely, the former becomes 
extended and the latter becomes localized. 

In Fig.~\ref{fig-states}(c), the dot sizes are proportional to the oscillator strength ($f_{if}$), defined as
\begin{equation}
f_{if}=\frac{2m^{*}}{\hbar^{2}}(E_{i}-E_{f})\left|\langle\Psi_{i}(x)|x|\Psi_{f}(x)\rangle\right|^{2},\nonumber
\end{equation}
where $E_{i}$ and $E_{f}$ are the eigenvalues of the initial 
$\Psi_{i}(x)$ and final $\Psi_{f}(x)$ states of the transition, 
respectively. As we can observe, the localized state has a greater 
oscillator strength than the extended one. In the anticrossing we have 
a particular situation in which both the states can be excited, since 
they have reasonable oscillator strength, and they are sufficiently 
extended states (large $P$), allowing for current generation. 
Therefore the mixing of localized and extended states in the 
anticrossing regions are the origin to the photocurrent enhancement 
seen in Fig.~\ref{fig-IV}. The phenomenon could be related to the LZSM 
problem,\cite{Lan32,Zen32,Stuc32,Maj32,KAY} namely, a dynamical 
induced transition between states with energy levels in an 
anticrossing condition.\cite{SUN} In this picture, the electron can be 
easily driven out from the filter barrier region only when the 
localized-extended mixing of states occurs, leading to peaks in the 
photocurrent \textit{I(V)} curves. For semiconductor superlattices 
similar features are observed in the current-voltage characteristics 
and related to the resonant character of the tunneling.\cite{PET,AND} 
Different from previous works, the resulting effect in our system can 
be seen as the phenomenon of dynamic negative differential 
conductance, since Fig.~\ref{fig-IV} allows us to foresee a decrease 
in the (photo)current for specific electric field regions.

%%%%%%%%%%%%%%%%%%%%%%%%%%%%%%%%%%%%
%%%%%%%%%%%%%%%%%%%%%%%%%%%%%%%%%%%%%

%%%%%%%%%%%%%%%%%%%%%%%%%%%%%%%%%%%%
\begin{figure}
\includegraphics[width=7.0cm]{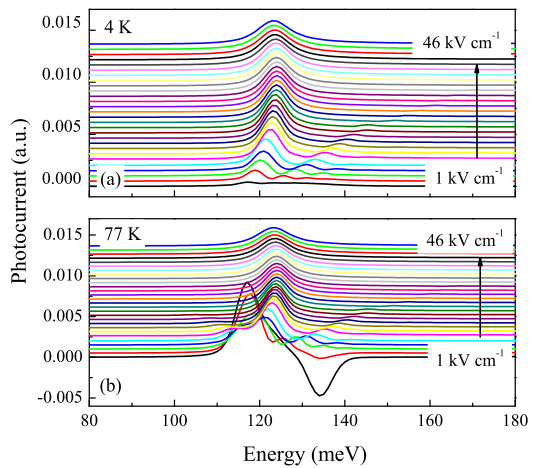}
\caption{\label{fig-boltzmann}(color online) Photocurrent spectra for different electric fields, ranging from 1 to 46 kV cm$^{-1}$, using the Boltzmann factor to simulate temperature effects for (a) 4 K and (b) 77 K.}
\end{figure}
%%%%%%%%%%%%%%%%%%%%%%%%%%%%%%%%%%%%%
  
Finally, as our method does not take in account temperature effects, 
we use a multiplicative Boltzmann factor in the calculation of the 
photocurrent to give us a qualitative understanding of the temperature 
influence on the MQW system. We stress that there is no doping in our 
model system, yet there will be a source to charge carriers in a real 
system. The effect of the temperature considered here concerns only 
the redistribution of these carriers throughout the energy levels. The 
photocurrent will thus be given by
\begin{equation}
I(T) \propto \sum_{QW} e^{-\frac{E_{QW}}{k_{B}T}}I_{QW},
\label{boltz}
\end{equation}
where the sum runs over all the quantum wells, $k_{B}$ is the Boltzmann constant,
$T$ is the temperature and
$I_{QW}$ is the photocurrent contribution calculated from a
single QW. The energies $E_{QW}$ are measured allways from the
lowest energy present in the structure, which in the case illustrated
in the inset of Fig.(\ref{fig-TotalCur}), corresponds to the energy
of the
$1^{st}$ QW on the right, whose energy is taken to be 
zero.

Fig.~\ref{fig-boltzmann}(a) shows the temperature dependent
photocurrent $I(T)$ obtained with this approach as a function
of the photon energy, for static electric 
fields varying from 1 to 46 kVcm$^{-1}$ and $T =4$ K.
We note that the behavior of the $T-$dependent
photocurrent is 
almost the same as the photocurrent calculated from the single
QW (the $1^{st}$ on the right) shown in the inset of Fig.~\ref{fig-TotalCur}
and whose results are depicted in Fig.~\ref{fig-PCbyQW}(a).
This can be undertood from the exponential Boltzmann
occupation factor that, for lower temperatures, tend to be negligible
for states other than the lowest energy one, which is also
more localized around its corresponding QW.

As the temperature is increased, the more energetic and delocalized levels 
become occupied. Fig.~\ref{fig-boltzmann}(b) shows the photocurrent spectra 
at $T$= 77 K, a temperature that yields contributions from electrons
that occupy the entire
structure. For low biases the spectra are quite similar to the zero 
temperature limit. However by increasing the electric field the Stark 
shift triggers the effect of the Boltzmann factor, and as a result the 
negative-photoconductance phenomenon effect discussed earlier is 
quenched.

\section{Conclusions}

In summary, we study the influence of an electric field provided by an 
external bias on the photocurrent generated in a semiconductor 
heterostructure that consists of multiple quantum wells with filter barriers.

Generally 
speaking, the main photocurrent peak is weakly Stark shifted since 
this peak is dominated by the transition between the bound state of a 
single quantum well (QW) and states in a
miniband-like group formed in the
lowest energy region of the continuum, in the same
position of the QW.
Peaks that are strongly shifted are related to transitions between a
QW bound state and miniband states of neighboring QWs. These 
peaks have intensities strongly dependent on the bias electric field
due to  spatial overlapping of the wave functions induced by the Wannier-Stark 
localization.

More importantly, we find a negative photoconductance, 
i.e., a decrease of the photocurrent with increasing the electric 
field, whose origin is in the state mixing of localized and extended 
states. The former are excited states belonging to the miniband 
created by the filter barriers. The 
mixing is very sensitive to the bias and the anticrossings of energy 
levels lead
to Landau-Zener-St¨uckelberg-Majorana\cite{Lan32,Zen32,Stuc32,Maj32}
transitions to extended states.
Reference~\onlinecite{SIB92} has experimentally found negative 
differential velocities in biased superlattices, however the
observed resonances are claimed to be related to interband
excitonic processes. In contrast, our calculations
suggest the possibility for observation of negative
differential conductance that is the result of localized-extended
mixing of intraband electronic states, wthin the conduction band.

Finally, in order to 
simulate temperature effects we averaged the photocurrent
over the quantum wells using a Boltzmann factor, with the
expected result
that lower energy quantum wells become increasingly more important
for the net
photocurrent signal as the temperature is lowered.
We believe this simple way of including
a Boltzmann factor and averaging over the quantum wells
give us a qualitative estimate for the temperature behavior
of the photocurrents.
 
\begin{acknowledgments}
The authors  are funded by DISSE-Instituto Nacional de Ci\^encia e Tecnologia de Nanodispositivos Semicondutores and Conselho Nacional de Desenvolvimento Cient\'ifico e Tecnol\'ogico (CNPq). MHD and MZM acknowledge financial support from Funda\c{c}\~ao de Amparo \`a Pesquisa do Estado de S\~ao Paulo (FAPESP).
\end{acknowledgments}

\end{document}